\documentclass[10pt,letterpaper]{article}
\usepackage[top=0.85in,left=2.75in,footskip=0.75in]{geometry}

\usepackage{amsmath,amssymb}

\usepackage{changepage}

\usepackage[utf8x]{inputenc}

\usepackage{textcomp,marvosym}

\usepackage{cite}

\usepackage{nameref,hyperref}

\usepackage[right]{lineno}

\usepackage{microtype}
\DisableLigatures[f]{encoding = *, family = * }

\usepackage[table]{xcolor}

\usepackage{array}

\newcolumntype{+}{!{\vrule width 2pt}}

\newlength\savedwidth

\raggedright
\usepackage[aboveskip=1pt,labelfont=bf,labelsep=period,justification=raggedright,singlelinecheck=off]{caption}

\makeatletter
\renewcommand{\@biblabel}[1]{\quad#1.}
\makeatother

\usepackage{lastpage,fancyhdr,graphicx}
\usepackage{epstopdf}
\fancyheadoffset[L]{2.25in}
\fancyfootoffset[L]{2.25in}
\usepackage{graphicx}
\usepackage{wasysym}
\usepackage{float}

\begin{document}
\vspace*{0.2in}

\begin{flushleft}
{\Large
\textbf\newline{Relationship between incidence of breathing obstruction and degree of muzzle shortness in pedigree dogs} % Please use "sentence case" for title and headings (capitalize only the first word in a title (or heading), the first word in a subtitle (or subheading), and any proper nouns).
}
\newline
\\
Richard~D.~Gill
\\
19 September 2022
\\
\bigskip
Mathematical Institute, Leiden University, Leiden, Netherlands\\
\url{https://www.math.leidenuniv.nl/~gill}\\

\bigskip

% Insert additional author notes using the symbols described below. Insert symbol callouts after author names as necessary.
% 
% Remove or comment out the author notes below if they aren't used.
%
% Primary Equal Contribution Note
% \Yinyang These authors contributed equally to this work.

% Additional Equal Contribution Note
% Also use this double-dagger symbol for special authorship notes, such as senior authorship.
% \ddag These authors also contributed equally to this work.

% \textcurrency Current Address: Dept/Program/Center, Institution Name, City, State, Country % change symbol to "\textcurrency a" if more than one current address note
% \textcurrency b Insert second current address 
% \textcurrency c Insert third current address

% \dag Deceased

% Group/Consortium Author Note
% \textpilcrow Membership list can be found in the Acknowledgments section.

% Use the asterisk to denote corresponding authorship and provide email address in note below.
email: gill@math.leidenuniv.nl

\end{flushleft}
\section*{Abstract}
There has been much concern about health issues associated with the breeding of short-muzzled pedigree dogs. The Dutch government commissioned a scientific report \emph{Fokken met Kortsnuitige Honden} (Breeding of short muzzled dogs), van Hagen (2019), and based on it rather stringent legislation, restricting breeding primarily on the basis of a single simple measurement of brachycephaly, the CFR: cranial-facial ratio. Van Hagen's work is a literature study and it draws heavily on statistical results obtained in three publications: Njikam (2009), Packer et al.~(2015), and Liu et al.~(2017). In this paper I discuss some serious shortcomings of those three studies and in particular show that Packer et al.\ have drawn unwarranted conclusions from their study. In fact, new analyses using their data leads to an entirely different conclusion.

% Please keep the Author Summary between 150 and 200 words
% Use first person. PLOS ONE authors please skip this step. 
% Author Summary not valid for PLOS ONE submissions.   

% \linenumbers

% Use "Eq" instead of "Equation" for equation citations.
\section*{Introduction}
The present work was commissioned by “Stichting Ras en Recht” (SRR; \emph{Foundation Justice for Pedigree dogs}) and focusses on the statistical research results of earlier papers summarized in the literature study \emph{Fokken met Kortsnuitige Honden} (Breeding of short-muzzled – brachycephalic -- dogs) by dr M.~van Hagen (2019). That report is the final outcome of a study commissioned by the Netherlands \emph{Ministry of Agriculture, Nature, and Food Quality}. It was used by the ministry to justify legislation restricting breeding of animals with extreme brachycephaly as measured by a low CFR, cranial-facial ratio.

An important part of van Hagen’s report is based on statistical analyses in three key papers: Njikam et al.~(2009), Packer et al.~(2015), and Liu et al.~(2017).  Notice: the paper Packer et al.~(2015) reports results from two separate studies, called by the authors Study 1 and Study 2. The data analysed in Packer et al.~(2015) study 1 was previously collected and analysed for other purposes in an earlier paper Packer et al.~(2013) which does not need to be discussed here.

In this paper I will focus on these statistical issues. My conclusion is the cited papers have many serious statistical shortcomings, which were not recognised by van Hagen (2019). In fact, a reanalysis of the Study 2 data investigated in Packer et al.~(2015) leads to conclusions completely opposite to those drawn by Packer et al., and completely opposite to the conclusions drawn by van Hagen. I come to the conclusion that the Packer et al.\ study 2 badly needs updating with a much larger replication study.

A very important question is just how generalisable are the results of those papers. There is no word on that issue in van Hagen (2019). I will start with discussing the paper which is most relevant to our question: Packer et al.~(2015).

An important prepartaory remark should be made concerning the term ``BOAS'', brachycephalic obstructive airway syndrome. It is a syndrome, which means: a name for some associated characteristics. ``Obstructed airways'' means: difficulty in breathing. ``Brachycephalic'' means: having a (relatively) short muzzle. Having difficulty in breathing is a symptom sometimes caused by having obstructed airways; it is certainly the case that the medical condition is often associated with having a short muzzle. That does not mean that having the short muzzle causes the medical condition. In the past, dog breeders have selected dogs with a view to accentuating certain features, such as a short muzzle: unfortunately, at the same time, they have sometimes selected dogs with other, less favourable characteristics at the same time. The two features of dogs' anatomies are associated, but one is not the cause of the other. ``BOAS'' really means: having obstructed airways \emph{and} a short muzzle.

\section*{Packer et al.~(2015): an exploratory and flawed paper}
Packer et al.~(2015) reports findings from \emph{two} studies. The sample for the first study, ``Study 1'', 700 animals, consisted of almost all dogs referred to the \emph{Royal Veterinary College Small Animal Referral Hospital} (RVC-SAH) in a certain period in 2012. Exclusions were based on a small list of sensible criteria such as the dog being too sick to be moved or too aggressive to be handled. However, this is not the end of the story. In the next stage, those dogs who actually were diagnosed to have BOAS (brachycephalic obstructive airway syndrome) were singled out, together with all dogs whose owners reported respiratory difficulties, except when such difficulties could be explained by respiratory or cardiac disorders. This resulted in a small group of only 70 dogs considered by the researchers to have BOAS, and it involved dogs of 12 breeds only. Finally, all the other dogs of those breeds were added to the 70, ending up with 152 dogs of 13 (!) breeds. (The paper contains many other instances of carelessness).

To return to Packer et al.~(2015) Study 1, this sample is a sample of dogs with health problems so serious that they are referred to a specialist veterinary hospital. One might find a relation between BOAS and CFR (craniofacial ratio) in that special population which is not the same as the relation in general. Moreover, the overall risk of BOAS in this special population is by its construction higher than in general. Breeders of pedigree dogs generally exclude already sick dogs from their breeding programmes.

This first study is justly characterised by the authors as exploratory. They had originally used the big sample of 700 dogs for a quite different investigation, Packer et al.~(2013). It is exploratory in the sense that they investigated a number of possible risk factors for BOAS besides CFR, and actually used the study to choose CFR as appearing to be the most influential risk factor, when each is taken on its own, according to a certain statistical analysis method, in which already a large number of prior assumptions had been built in. As I will repeat a few more times, the sample is too small to check those assumptions. I do not know if they also tried various simple transformations of the risk factors. Who knows, maybe the logarithm of a different variable would have done better than CFR.

In the second study (“Study 2”), they sampled anew, this time recruiting animals directly mainly from breeders but also from general practice. A critical selection criterium was a CFR smaller than 0.5, that number being the biggest CFR of a dog with BOAS from Study 1. They specially targeted breeders of breeds with low CFR, especially those which had been poorly represented in the first study. Apparently, the Affenpinscher and Griffon Bruxellois are not often so sick that they get referred to the RVC-SAH; of the 700 dogs entering Study 1 there was, for instance, just 1 Affenpinscher and only 2 Griffon Bruxellois. Of course, these are also relatively rare breeds. Anyway, in Study 2, those numbers became 31 and 20. So: the second study population is not so badly biased towards sick animals as the first. Unfortunately, the sample is much, much smaller, and per breed, very small indeed, despite the augmentation of rarer breeds.

\begin{figure}[H]
\includegraphics[width=\textwidth]{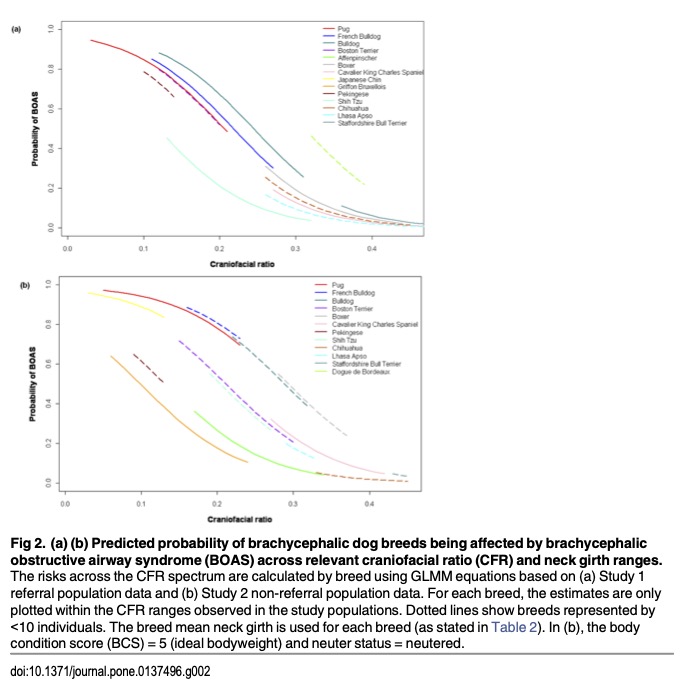}
\caption{\bf Figure 2, and its original caption, from Packer et al.~(2015).}
\small Predicted probability of brachycephalic dog breeds being affected by brachycephalic obstructive airway syndrome (BOAS) across relevant craniofacial ratio (CFR) and neck girth ranges. The risks across the CFR spectrum are calculated by breed using GLMM equations based on (a) Study 1 referral population data and (b) Study 2 non-referral population data. For each breed, the estimates are only plotted within the CFR ranges observed in the study populations. Dotted lines show breeds represented by $<$10 individuals. The breed mean neck girth is used for each breed (as stated in Table 2). In (b), the body condition score (BCS) = 5 (ideal bodyweight) and neuter status = neutered.
\label{fig1}
\end{figure}

Now it is important to turn to technical comments concerning what perhaps seems to speak most clearly to the non-statistically schooled reader, namely, Figure 2 of Packer et al., which I reproduce here, together with the figure's original caption.

In the abstract of their paper, they write ``we show […] that BOAS risk increases sharply in a non-linear manner''. They do no such thing! They \emph{assume} that the log odds of BOAS risk $p$, that is: $\log(p/(1-p))$,  \emph{depends exactly linearly on CFR and moreover with the same slope for all breeds}. The small size of these studies forced them to make such an assumption. It is a conventional “convenience” assumption. Indeed, this is an exploratory analysis, moreover, the authors' declared aim was to come up with a single risk factor for BOAS. They were forced to extrapolate from breeds which are represented in larger numbers to breeds of which they had seen many less animals. They use the whole sample to estimate just one number, namely the slope of $\log(p/(1-p))$ as an assumed linear function of CFR. Each small group of animals of each breed then moves that linear function up or down, which corresponds to moving the curves to the right or to the left. Those are not \emph{findings} of the paper. They are conventional model \emph{assumptions} imposed by the authors from the start for statistical convenience and statistical necessity and completely in tune with their motivations.

One indeed sees in the graphs that all those beautiful curves are essentially segments of the same curve, shifted horizontally. This has not been shown in the paper to be true. It was \emph{assumed} by the authors of the paper to be true. Apparently, that assumption worked better for CFR than for the other possible criteria which they considered: that was demonstrated by the exploratory (the authors own characterisation!) Study 1. When one goes from Study 1 to Study 2, the curves shift a bit: it is definitely a different population now. 

There are strange features in the colour codes. Breeds which should be there are missing, breeds which shouldn’t be there are. The authors have exchanged graphs (a) and (b)! This can be seen by comparing the minimum and maximum predicted risks from their Table 2.

Notice that these curves represent predictions for neutered dogs with breed mean neck girth,  breed ideal body condition score (breed ideal body weight).  I don't know whose definition of ideal is being used here. The graphs are not graphs of probabilities for dog breeds, but model predictions for particular classes of dogs of various breeds. They depend strongly on whether or not the model assumptions are correct. The authors did not (and could not) check the model assumptions: the sample sizes are much too small. 

By the way, breeders' dogs are generally not neutered. Still, one third of the dogs in the sample were neutered, so the ``baseline'' does represent a lot of animals. Notice that there is no indication whatsoever of statistical uncertainty in those graphics. The authors apparently did not find it necessary to add error bars or confidence bands to their plots. Had they done so, the pictures would have given a very, very different impression.

In their discussion, the authors write ``Our results confirm that brachycephaly is a risk factor for BOAS and for the first time quantitatively demonstrate that more extreme brachycephalic conformations are at higher risk of BOAS than more moderate morphologies; BOAS risk increases sharply in a non-linear manner as relative muzzle length shortens''. I disagree strongly with their appraisal. The vaunted non-linearity was just a conventional and convenience (untested) assumption of linearity in the much more sensible log-odds scale. They did not test this assumption and most importantly, they did not test whether it held for each breed considered separately. They could not do that, because both of their studies were much, much too small. Notice that they themselves write, ``we found some exceptional individuals that were unaffected by BOAS despite extreme brachycephaly'' and it is clear that these exceptions were found in specific breeds. But they do not tell us which.

They also tell us that other predictors are important next to CFR. Once CFR and breed have been taken into account (in the way that they take it into account!), neck girth (NG) becomes very important.

They also write, ``if society wanted to eliminate BOAS from the domestic dog population entirely then based on these data a quantitative limit of CFR no less than 0.5 would need to be imposed''. They point out that it is unlikely that society would accept this, and moreover, it would destroy many breeds which do not have problems with BOAS at all! They mention, ``several approaches could be used towards breeding towards more moderate, lower-risk morphologies, each of which may have strengths and weaknesses and may be differentially supported by stakeholders involved in this issue''.

This paper definitely does not support imposing a single simple criterion for all dog breeds, much as its authors might have initially hoped that CFR could supply such a criterion.

In a separate section I will test their model assumptions, and investigate the statistical reliability of their findings.

\section*{Liu et al.~(2017): an excellent study, but of only three breeds}
Now I turn to the other key paper, Liu et al.~(2017). On this 8 author paper, the last and senior author, Jane Ladlow, is a very well known authority in the field. This paper is based on a study involving 604 dogs of only three breeds, and those are the three breeds which are already known to be most severely affected by BOAS: bulldogs, French bulldogs, and pugs. They use similar statistical methodology to Packer et al., but now they allow each breed to have a different shaped dependence on CFR. Interestingly, the effects of CFR on BOAS risk for pugs, bulldogs and French bulldogs are not statistically significant. Whether or not they are the same across those three breeds becomes, from the statistical point of view, an academic question.

The statistical competence and sophistication of this group of authors can be seen at a glance to be immeasurably higher than that of the group of authors of Packer et al. They do include indications of statistical uncertainty in their graphical illustrations. They state, ``in our study with large numbers of dogs of the three breeds, we obtained supportive data on NGR (neck girth ratio: neck girth / chest girth), but only a weak association of BOAS status with CFR in a single breed.'' Of course, part of that could be due to the fact that, in their study, CFR did not vary much within each of those three breeds, as they themselves point out.  I did not yet re-analyse their data to check this. CFR was certainly highly variable in these three breeds in both of Packer et al.'s studies, see the figures above, and again in Liu et al.\ as is apparent from my Figure 2 below. But Liu et al.\  also point out that anyway, ``anatomically, the CFR measurement cannot determine the main internal BOAS lesions along the upper airway''.

Another of their concluding remarks is the rather interesting ``overall, the conformational and external factors as measured here contribute less than 50\% of the variance that is seen in BOAS''. In other words, BOAS is not very well predicted by these shape factors. They conclude, ``breeding \emph{toward} [my emphasis] extreme brachycephalic features should be strictly avoided''. I should hope that nowadays, no recognised breeders deliberately try to make known risk features even more pronounced.

Liu et al.\ studied only bulldogs, French bulldogs and pugs. The CFR's of these breeds do show within breed statistical variation. The study showed that a different anatomical measure was an excellent predictor of BOAS. Liu et al.\ moreover explain anatomically and medically why one should not expect CFR to be relevant for the health problems of those races of dogs.

It is absolutely not true that almost all of the animals in that study have BOAS. The study does not investigate BOS. The study was set up in order to investigate the exploratory findings and hypotheses of Packer et al.\ and it rejects them, as far as the three races they considered were concerned. Packer et al.\ hoped to find a simple relationship between CFR and BOAS for all brachycephalic dogs but their two studies are both much too small to verify their assumptions. Liu et al.\ show that for the three races studied, the relation between measurements of body structure and ill health associated with them, varies between races.

\begin{figure}[H]
\includegraphics[width=\textwidth]{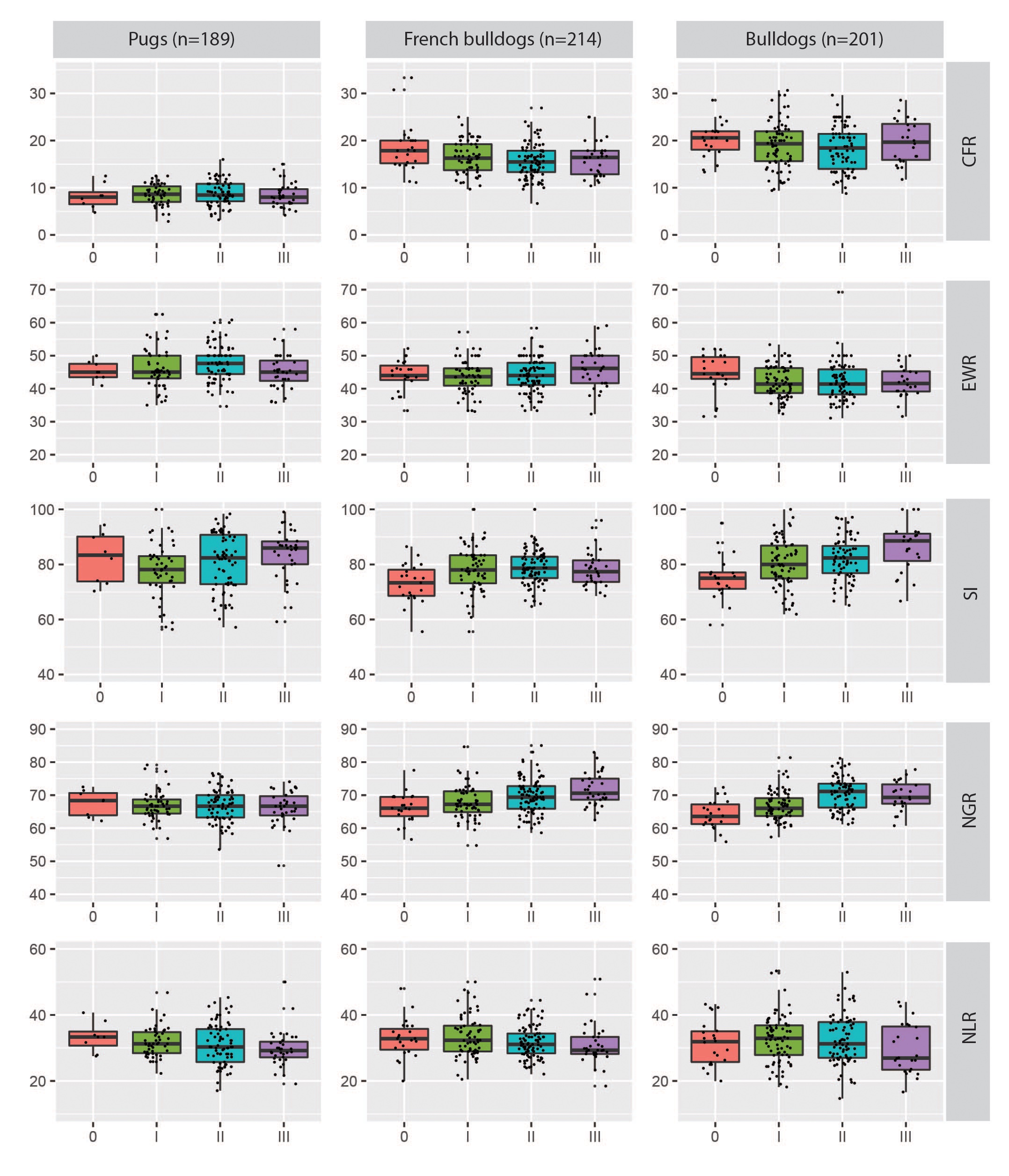}
\caption{Supplementary materical \textbf{S1 Fig} from Liu et al.~(2017.)}
\small Boxplots show the distribution of the five conformation ratios against BOAS functional grades.
The x-axis is BOAS functional grade; the y-axis is the ratios in percentage. CFR, craniofacial ratio; EWR, eye with ratio; SI, skull index; NGR, neck girth ratio; NLR, neck length ratio.
\label{fig2}
\end{figure}

In contradiction to the opinion of van Hagen (2019), there are no ``contradictions'' between the studies of Packer et al.\ and Liu et al. The first comes up with some guesses, based on tiny samples from each breed. The second investigates those guesses but discovers that they are wrong for the three races most afflicted with BOAS. Study 1 of Packer et al.\ is a study of sick animals, but Study 2 is a study of animals from the general population. Liu et al.\ is a study of animals from the general population. (To complicate matters, Njikam et al., Packer et al. and Liu et al.\ all use slightly different definitions or categorisations of BOAS.)

Njikam et al.\ (2009), like the later researchers in the field, fit logistic regression models. They exhibit various associations between illness and risk factors per breed. They do not quantify brachycephaly by CFR but by a similar measure, BRA, the ratio of width to length of skull. CFR and BRA are approximately non-linear one-to-one functions of one another (this would be exact if skull length equalled skull width plus muzzle length, i.e., assuming a spherical cranium), so a threshold criterium in terms of one can be roughly translated into a threshold criterium in terms of the other. Their samples are again, unfortunately, very small (the title of their paper is very misleading).

Their main interest is in genetic factors associated with BOAS apart from the genetic factors behind CFR, and indeed they find such factors! In other words, this study shows that BOAS is very complex. Its causes are multifactorial. They have no data at all on the breeds of primary interest to SRR: these breeds are not much afflicted by BOAS! It seems that van Hagen again has a reading of Njikam et al.\ which is not justified by that paper's content.

\section*{Packer et al. (2015) Study. 2, revisited}

Fortunately, the data-sets used by the publications in \emph{PLoS ONE} are available as ``supplementary material'' on the journal's webpages. First of all I would like to show a rather simple statistical graphic which shows that the relation between BOAS and CFR in Packer et al.'s Study 2 data does not look at all as the authors hypothesized. First, here are the numbers: a table of numbers of animals with and without BOAS in groups split according to CFR as percentage, in steps of 5\%. The authors recruited animals mainly from breeders, with CFR less than 50\%. It seems there were none in their sample with a CFR between 45\% and 50\%.
\begin{table}[H]
\centering
\begin{tabular}{rrrrrrrrrr}
  \hline
BOAS & (0,5] & (5,10] & (10,15] & (15,20] & (20,25] & (25,30] & (30,35] & (35,40] & (40,45] \\ 
  \hline
0 &   1 &   4 &  12 &  12 &  22 &  13 &  12 &   4 &  15 \\ 
  1 &   9 &  11 &  19 &   5 &   5 &   4 &   1 &   2 &   3 \\ 
   \hline
\end{tabular}
\caption{BOAS versus CFR group} 
\end{table}

This next figure is a simple ``pyramid plot'' of percentages with and without BOAS per CFR group.
I am not taking account of breed of these dogs, nor of other possibly explanatory factors. However, as we will see, the suggestion given by the plot seems to be confirmed by more sophisticated analyses. And that suggestion is: BOAS has a roughly constant incidence of about 20\% among dogs with a CFR between 20\% and 45\%. Below that level, BOAS incidence increases more or less linearly as CFR further decreases.

Be aware that the sample sizes on which these percentages are based are very, very small. 

\begin{figure}[H]
\includegraphics[width=\textwidth]{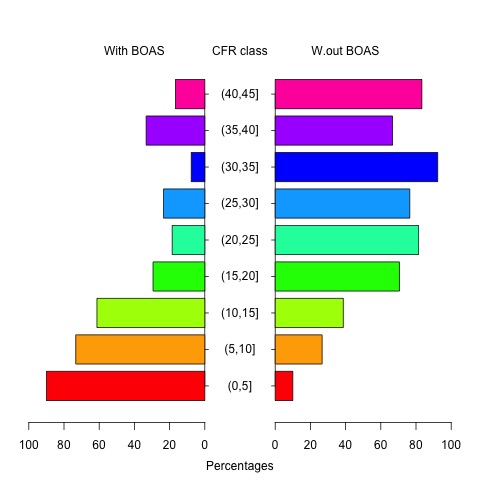}
\caption{Pyramid plot, data from Packer et al. Study 2}
\label{fig3}
\end{figure}

Could it be that the pattern shown in Figure 3 is caused by other important characteristics of the dogs, in particular breed? In order to investigate this question, I first of all fitted a logistic regression model with only CFR, and then a \emph{smooth} logistic regression model with only CFR. In the latter, the effect of CFR on BOAS is allowed to be any smooth function of CFR -- not a function of particular shape. The two fitted curves are seen in Figure 4. The solid line is the smooth, the dashed line is the standard logistic curve.

\begin{figure}[H]
\includegraphics[width=\textwidth]{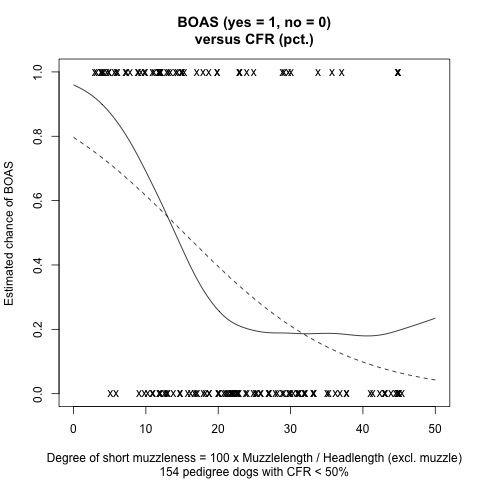}
\label{fig4}
\end{figure}

This analysis confirms the impression of the pyramid plot. However, the next results which I obtained were dramatic. I added to the smooth model also Breed and Neutered-status, and also investigated some of the other variables which turned up in the papers I have cited. It turned out that \emph{``Breed'' is not a useful explanatory factor}. CFR is hardly significant. Possibly, just one particular breed is important: the Pug. The differences between the others are negligable (once we have taken account of CFR). The variable ``neutered'' remains as somewhat important.

Here is the best model which I found. As far as I can see, the Pug is a rather different animal from all the others. On the logistic scale, even taking account of CFR, Neckgirth and Neuter status, being a Pug increases the log odds ratio for BOAS by 2.5. Below a CFR of 20\%, each 5\% decrease in CFR increases the log odds ratio for BOAS by 1, so is associated with an increase in incidence by a factor of close to 3. In the appendix can be seen what happens when we allow each breed to have its own effect. We can no longer separate the influence of Breed from CFR and we cannot say anything about any individual breeds, except for one.
\newpage

\begin{table}[H]
\begin{center}
\begin{tabular}{l c c}
\hline
 & Model 1 \\
\hline
(Intercept)                     & $-3.86^{***}$ 
                                & $(0.97)$      \\
(CFRpct - 20) * (CFRpct $<$ 20) & $-0.20^{***}$ 
                                & $(0.05)$      \\
Breed == ``Pug":TRUE              & $2.48^{***}$  
                                & $(0.71)$      \\
NECKGIRTH                       & $0.06^{*}$    
                                & $(0.03)$      \\
NEUTER:Neutered                  & $1.00^{*}$    
                                & $(0.50)$      \\
\hline
AIC                             & $144.19$      \\
BIC                             & $159.37$      \\
Log Likelihood                  & $-67.09$      \\
Deviance                        & $134.19$      \\
Num. obs.                       & $154$         \\
\hline
\multicolumn{2}{l}{\scriptsize{$^{***}p<0.001$; $^{**}p<0.01$; $^{*}p<0.05$}}
\end{tabular}
\caption{A very simple model (GLM, logistic regression)}
\label{table:coefficients}
\end{center}
\end{table}

The pug is in a bad way. But we knew that before. Packer Study 2 data:
\begin{table}[H]
\centering
\begin{tabular}{rrr}
  \hline
  & W.out BOAS & With BOAS \\ 
  \hline
Not Pug &  92 &  30 \\ 
  Pug &   3 &  29 \\ 
   \hline
\end{tabular}
\end{table}

The graphs of Packer et al.~in Figure 1 are a fantasy. Reanalysis of their data shows that their model assumptions are wrong. We already knew that BOAS incidence, Breed, and CFR are closely related and naturally they see that again in their data. But the actual possibly Breed-wise relation between CFR and BOAS is completely different from what their fitted model suggests. In fact, the relation between CFR and BOAS seems to be much the same for all breeds, except possibly for the Pug.

\section*{Final remarks}

The paper Packer et al.~(2015) is rightly described by its authors as exploratory. This means: it generates suggestions for further research. On the other hand, the later paper by Liu et al.~(2017) is an excellent piece of research. It follows up the suggestions of Packer et al. But it does not find confirmation of their hypotheses. On the contrary, it gives strong evidence that they were false. Unfortunately, it only studies three breeds, and those breeds are breeds where we already know action should be taken. But already on the basis of a study of just those three breeds, it comes out strongly against taking one single simple criterion, the same for all breeds, as basis for legislation on breeding.

Further research based on reanalysis of the data of Packer et al.~(2015) shows that the main assumptions of those authors were wrong and that, had they made more reasonable assumptions, completely different conclusions would have been drawn from their study.

The conclusion to be drawn from the works I have discussed is that it is unreasonable to suppose that a single simple criterion, the same for all breeds, can be a sound basis for legislation on breeding. Packer et al.~clearly hoped to find support for this, but failed: Liu et al.~scuppered that dream. Reanalysis of their data with more sophisticated statistical tools shows that they should already have seen that they were betting on a wrong horse.

Below a CFR of 20\%, further decrease in CFR is associated with higher incidence of BOAS. There is not enough data on every breed to see if this relation is the same for all breeds. For Pugs, things are much worse. For some breeds, it might not be so bad.

Study 2 of Packer et al.~(2015) needs to be replicated, with much larger sample sizes.

\nolinenumbers

% Either type in your references using
% \begin{thebibliography}{}
% Text
% \end{thebibliography}
%
% or
%
% Compile your BiBTeX database using our plos2015.bst
% style file and paste the contents of your .bbl file
% here. See http://journals.plos.org/plosone/s/latex for 

\section*{Appendix: what happens when we try to separate``Breed" from ``CFR"}

\begin{table}[H]
\begin{center}
\begin{tabular}{l c c}
\hline
 & Model 2 \\
\hline
(Intercept)                           & $-3.73^{*}$    
                                      & $(1.65)$        \\
Breed:American Bulldog                 & $-43.51$        
                                      & $(67108864.00)$ \\
Breed:Bolognese                        & $-40.45$       
                                      & $(47453132.81)$ \\
Breed:Boston Terrier                   & $0.35$          
                                      & $(1.84)$        \\
Breed:Boxer                            & $1.23$          
                                      & $(1.72)$        \\
Breed:Bulldog                          & $1.04$          
                                      & $(1.68)$        \\
Breed:Cavalier King Charles Spaniel    & $0.82$          
                                      & $(1.37)$        \\
Breed:Chihuahua                        & $-42.77$       
                                      & $(38745320.70)$ \\
Breed:Dogue de Bordeaux                & $-43.35$        
                                      & $(67108864.00)$ \\
Breed:French Bulldog                   & $2.36$          
                                      & $(1.59)$        \\
Breed:Griffon Bruxellois               & $-0.97$         
                                      & $(1.18)$        \\
Breed:Japanese Chin                    & $1.70$          
                                      & $(1.46)$        \\
Breed:Lhasa Apso                       & $1.75$          
                                      & $(1.63)$        \\
Breed:Mastiff cross                    & $1.97$          
                                      & $(2.60)$        \\
Breed:Pekingese                        & $-45.60$        
                                      & $(38745320.70)$ \\
Breed:Pug                              & $2.69^{*}$      
                                      & $(1.26)$        \\
Breed:Pug cross                        & $-44.79$        
                                      & $(47453132.81)$ \\
Breed:Rottweiler                       & $-43.29$        
                                      & $(47453132.81)$ \\
Breed:Shih Tzu                         & $0.16$          
                                      & $(1.23)$        \\
Breed:Staffordshire Bull Terrier       & $-43.37$        
                                      & $(47453132.81)$ \\
Breed:Staffordshire Bull Terrier Cross & $2.36$          
                                      & $(2.07)$        \\
Breed:Tibetan Spaniel                  & $-44.14$       
                                      & $(67108864.00)$ \\
Breed:Victorian Bulldog                & $-43.16$        
                                      & $(67108864.00)$ \\
NECKGIRTH                             & $0.06$          
                                      & $(0.06)$        \\
NEUTER:Neutered                        & $1.80^{*}$      
                                      & $(0.84)$        \\
EDF: s(CFRpct)                        & $1.00$          
                                      & $(1.00)$        \\
\hline
AIC                                   & $158.59$        \\
BIC                                   & $237.55$        \\
Log Likelihood                        & $-53.29$        \\
Deviance                              & $106.59$        \\
Deviance explained                    & $0.48$          \\
Dispersion                            & $1.00$          \\
R$^2$                                 & $0.46$          \\
GCV score                             & $0.03$          \\
Num. obs.                             & $154$           \\
Num. smooth terms                     & $1$             \\
\hline
\multicolumn{2}{l}{\scriptsize{$^{***}p<0.001$; $^{**}p<0.01$; $^{*}p<0.05$}}
\end{tabular}
\caption{ A more complex model (GAM, logistic regression)}
\label{table:coefficients2}
\end{center}
\end{table}

That model allowing each breed to have its own separate ``fixed'' effect was not a success. That certainly was presumably the motivation to make ``Breed'' a random, not a fixed, effect in the Packer et al.\ publication, because treating breed effects as drawn from a normal distribution and assuming the same effect of CFR for all breeds disguises the multicollinearity and lack of information in the data. Many breeds, most of them contributing only one or two animals, enabled the authors' statistical software to compute an overall estimate of ``variability between breeds'' but the result is pretty meaningless.

Further inspection shows that many breeds are only represented by  1or 2 animals in the study. Only five are in something a bit like reasonable numbers. These five are the Affenpinscher, Cavalier King Charles Spaniel, Griffon Bruxellois, Japanese Chin and Pug; in numbers  31, 11, 20, 10, 32. I fitted a GLM (logistic regression) trying to explain BOAS in these 105 animals their breed together with variables CFR, BCR, and so on. Still then, the multicollinearity between all these variables is so strong that the best model did not include CFR at all. In fact: once BCS (Body Condition Score) was included, no other variable could be added without almost everything becoming statistically insignificant. Not surprisingly, it is good to have a good BCS. Being a Pug or a Japanese Chin is disastrous. Cavalier King Charles Spaniel is intermediate. Affenpinscher and Griffon Bruxellois have the least BOAS (and about the same amount, namely an incidence of 10\%), even though the mean CFR's of these two species seem somewhat different (0.25, 0.15).

Had the authors presented p-values and error bars the paper would probably never have been published. The study should be repeated with a sample 10 times larger.

\section*{Acknowledgments}
This work was partly funded by ``Stichting Ras en Recht'' (SRR; Foundation Justice for Pedigree dogs). The author accepted the commission by SSR to review statistical aspects of AE van Hagen's report ``Breeding of short muzzled dogs'' under condition that he would report his honest professional and scientific opinion on van Hagen's literature study and its sources. 

\end{document}